# Focal-UNet: UNet-like Focal Modulation for Medical Image Segmentation

MohammadReza Naderi[*,1], MohammadHossein Givkashi[*,1], Fatemeh Piri[1], Nader Karimi[1], Shadrokh Samavi[1,2,3]

[1]Department of Electrical and Computer Engineering, Isfahan University of Technology, 84156-83111, Iran
[2]Department of Electrical and Computer Engineering, McMaster University, L8S 4L8, Canada
[3]Department of Computer Science, Seattle University, Seattle, 98122 USA

**Abstract**

Recently, many attempts have been made to construct a transformer base U-shaped architecture, and new methods have been proposed that outperformed CNN-based rivals. However, serious problems such as blockiness and cropped edges in predicted masks remain because of transformers' patch partitioning operations. In this work, we propose a new U-shaped architecture for medical image segmentation with the help of the newly introduced focal modulation mechanism. The proposed architecture has asymmetric depths for the encoder and decoder. Due to the ability of the focal module to aggregate local and global features, our model could simultaneously benefit the wide receptive field of transformers and local viewing of CNNs. This helps the proposed method balance the local and global feature usage to outperform one of the most powerful transformer-based U-shaped models called Swin-UNet. We achieved a 1.68% higher DICE score and a 0.89 better HD metric on the Synapse dataset. Also, with extremely limited data, we had a 4.25% higher DICE score on the NeoPolyp dataset. Our implementations are available at: https://github.com/givkashi/Focal-UNet

## 1      Introduction

Computer vision has revolutionized medical image analysis by benefiting from the deep learning concept. Image segmentation is a vital task of medical image analysis. Particularly, accurate and generalizable medical image segmentation can play a base role in computer-aided diagnosis and image-guided clinical surgery [1], [2]. U-shaped networks were the most famous architectures in medical image segmentation due to their low complexity and high performance. The down-sampling, up-sampling, and skip-connection concepts apply the U-shaped architectures to different image segmentation problems. In recent years, transformers [3], due to their modeling capacity, have been applied to multiple image processing tasks. Mainly, transformers have a larger receptive field than convolutional rivals, which helps them learn more generalizable features between input and ground truth images. Designing U-shaped architectures using transformer blocks have been studied on several medical image segmentation tasks. These architectures achieved better results than convolutional base methods. Therefore, in this work, we investigate the possibility of designing new U-shaped architecture using recently proposed focal modulation network blocks. In multiple aspects, these blocks have shown better classification and object detection results than one of the most famous transformer blocks, the Swin transformer [4]. Traditional transformers such as trans-UNet [5] and Swin-UNet [6] require pre-trained weights on a large dataset, and besides, their performances drop dramatically in the minimal training sets scenarios. The main contributions

---

of the proposed method could be considered below:
- Using focal module as a base block of U-shaped architecture
- Greatening encoder layers depths to increase problem modeling capacity of U-shaped architecture
- Sliming decoder layers depths to decrease the vanishing gradient problem of U-shaped architecture
- Comparing the efficiency of the proposed method and Swin-UNet on handling small medical dataset

The structure of this paper is as follows: In section 2, we will present some previous relevant methods. Then, section 3 offers the proposed method. Finally, experimental results are detailed in section 4, and the conclusion is presented in section 5.

## 2  Related Works

In subsection 2.1, we will examine existing medical image segmentation methods based on CNN. After that, in subsection 2.2, we will look at some medical image processing based on transformers. Finally, in subsection 3.3, we will explain some medical image segmentation methods based on transformers we will use in our proposed method.

### 2.1  Medical image segmentation based on CNN

Considering the performance of CNN in image processing tasks, most deep learning methods in medical image segmentation rely on CNN-based architectures, specifically, encoder-decoder architectures. U-net [1] is the first encoder-decoder CNN-based method demonstrating high image segmentation performance. The remarkable performance of the U-net resulted in many variants of U-shaped structures. For instance, in weighted Res-UNet [2], a weighted attention mechanism is used to segment small regions. U-net++ [7] is a U-shaped structure with a series of nested, dense skip pathways leading to less semantic gap between the feature maps of the encoder and decoder. U-net3+ [8] applies a full-scale skip connection to combine low-level and high-level details. Dense-UNet [9] introduced a hybrid densely connected UNet in order to optimize both intra-slice and inter-slice features. ENS-UNet [10] is another U-shaped architecture with no need for massive pre-processing and post-processing. C-UNet [11] takes advantage of Inception-like convolutional block, recurrent convolutional block, and dilated convolutional layers for skin lesion segmentation. In general, CNN-based methods are widely used in image segmentation tasks due to the ability to extract features effectively by paying attention to adjacent pixels.

### 2.2  Medical image processing based on transformers

Although CNN-based methods obtained superb results in medical image segmentation, these methods cannot model long-range dependencies due to the intrinsic locality of convolutional operation. After the success of transformers in Natural Language Processing (NLP) fields [12], a vision transformer (ViT) [3] has been proposed as a transformer in image processing tasks. Since transformers can capture long-range dependencies of input sequences using a multi-head self-attention (MSA) mechanism, ViT demonstrated high performance in image segmentation, image recognition, and object-detection tasks. DeiT [13] utilized ViT as its backbone and reduced the data dependency of ViT using data-efficient training strategies.

One of the drawbacks of ViT is its computational complexity which is quadratic to the image size, so using ViT for high-resolution images is challenging. In [4], a new network called the Swin transformer has been introduced to address the computational complexity issue. Instead of using the MSA mechanism, the Swin transformer exploits Window MSA (W-MSA) and a Shifted Window MSA (SW-MSA) mechanisms. Using these mechanisms reduces computational complexity from quadratic to linear. Also, because of hierarchical feature maps in the Swin transformer, it can efficiently be replaced by the ResNet backbone in many vision tasks. As a result, the Swin transformer earned state-of-the-art performance in image recognition and dense prediction tasks like object detection and semantic segmentation.

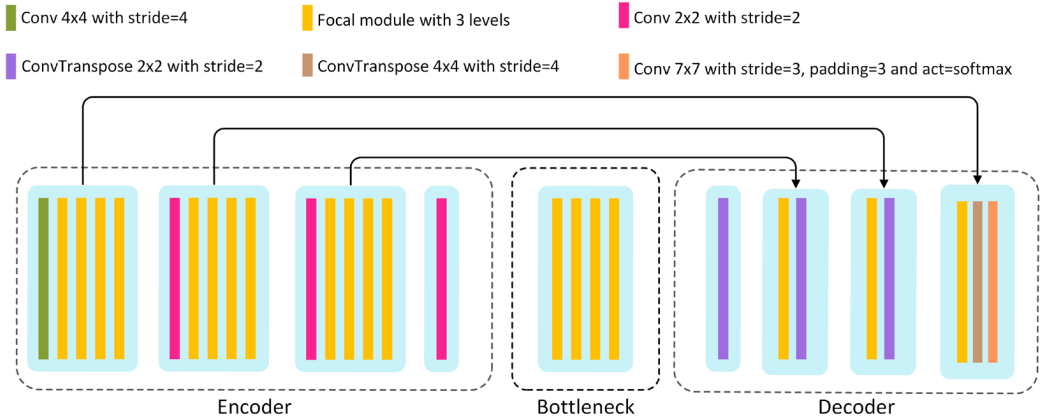

Figure 1: Overview of the Focal-UNet architecture.

The focal transformer [14] employed a focal self-attention mechanism incorporating fine-grained local and coarse-grained global interactions. So focal transformer utilized both short- and long-range dependencies. The focal self-attention mechanism has larger receptive fields with less time and memory cost than the standard self-attention mechanism. A focal modulation network (focal nets) [15] is proposed as a new structure that employs focal modulation instead of self-attention. In other words, focal nets have replaced self-attention or focal self-attention mechanisms with focal modulation. Focal modulation aggregates the input context in the first step, while in the self-attention mechanism, aggregation is done after interaction between the query and input context. Focal nets consist of three significant sections; first, the hierarchical contextualization part captures short to long-range dependencies for each input token in a hierarchical manner. Second, a gated aggregation section to aggregate context features based on the content of the input token. And the last section is a modulator that the output of the gated aggregation section is fed into it. Focal nets outperform state-of-the-art methods such as the Swin transformer and focal transformer in image classification, object detection, and semantic segmentation.

### 2.3   Medical image segmentation based on transformers

TransUNet [5] is the first structure that combines transformer and CNN in medical image segmentation tasks to use the advantages of both networks. Afterward, many other researchers attempted to apply a transformer in a CNN-based architecture for high-performance image segmentation. For instance, IT-UNet [16] combines CNN and transformer for organs-at-risk segmentation. FcTC-UNet [17] is a hybrid method based on CNN and transformers to thoracic segmentation. UCATR [18] is another example of this hybrid structure used for lesion segmentation. Medical transformer [19] and transfuse [20] are examples of utilizing both CNN and transformer. Despite the above models, some methods are purely transformer based to segment medical images. Swin-UNet [6] is an instance of this category. Swin-UNet is a U-shaped pure transformer-based structure that uses a hierarchical Swin transformer in both encoder and decoder for local and global semantic feature learning. Experimental results have shown that these convolutional free architectures can result in more accurate segmentation in some cases. In this work, we propose a pure transformer-based U-shaped structure that exploits focal modulation blocks [15] in both the encoder and decoder. Despite other medical image segmentation models that require a large dataset to achieve high performance, our proposed method can outperform state-of-the-art methods using small datasets.

# 3   Proposed Method

The architecture of our proposed method is illustrated in Figure 1. Focal-UNet has an encoder, bottleneck, decoder, and skip connection. We will explain each part separately.

**Encoder:** The encoder is constructed based on three focal layers. Each layer includes four Focal modulation blocks (FM blocks) and a down-sampling convolution block. FM block is adapted from [15]. Due to its larger receptive field for the network, it has a higher capacity to extract more generalizable features from the input feature vectors. A higher receptive field helps the network consider more generalizable features during the training, which seems effective for segmentation problems. Furthermore, this block is more computationally efficient because [15] used Depth Wise Convolution in its architecture. To summarize the above intuitions, this FM block seems prominent to use as a base block to construct a U-shaped architecture. The purpose of an additional convolution block in each layer is to reduce the image sizes and increase the number of channels, as is regular in designing U-shaped architectures.

**Bottleneck:** The bottleneck is constructed based on four FM blocks to extract deeper features from input and increase the capacity of the network to model more complex problem spaces.

**Decoder:** Decoder also is constructed similarly to the encoder part. The main difference is to use one FM block in each layer and a deconvolution block to up-sampling the features and decrease the number of channels in each stage by a factor of two.

We were inspired by [21] to increase the encoder capacity by using four FM blocks in each down-sampling layer of the encoder. As discussed in [21], having deeper encoder layers is more critical to extract meaningful features from the inputs, and having more shallower decoder layers help the network to backpropagate stronger gradients to the encoder part, which will prevent the gradient vanishing problem that concludes in better results.

**Skip connection:** One of the critical elements of the U-shaped architectures is the skip connection which leads the network to use higher and lower-level features simultaneously to construct the desired outputs. We also used a skip connection in the proposed architecture, similar to U-shaped models, to boost the feature transfer from the input to the output. The skip connections also boost the gradients that are backpropagated in the network.

# 4   Experimental Results

In this section, we investigate the effectiveness of Focal-UNet architecture by conducting experiments on Synapse and NeoPolyp datasets.

## 4.1   Datasets

**Synapse multi-organ segmentation dataset (Synapse):** The dataset contains 30 cases with 3779 axial abdominal clinical CT images. Following [5], [6], we used 18 samples as training data and the rest 12 samples as the test set. We used the average Dice-Similarity coefficient (DSC) and average Hausdorff Distance (HD) as evaluation metrics to evaluate our method on eight abdominal organs (aorta, gallbladder, spleen, left kidney, right kidney, liver, pancreas, spleen, and stomach).

**BKAI-IGH NeoPolyp-Small [22]:** This dataset includes 1200 colon polyps images with fine-grained segmentation annotations. The training set consists of 1000 images, and the test set consists of 200 images. All polyps are classified into neoplastic or non-neoplastic classes, denoted by red and green colors. This dataset is a part of a more extensive dataset called NeoPolyp. The ground truths of the test set of this dataset are not publicly available; therefore, only DSC is reported using the Kaggle submission mechanism.

Table 1: Comparing the performance of Focal-UNet with other methods on Synapse dataset.

| Methods | DSC↑ | HD↓ | Aorta | Gallbladder | Kidney (L) | Kidney (R) | Liver | Pancreas | Spleen | Stomach |
|---|---|---|---|---|---|---|---|---|---|---|
| V-Net [25] | 68.81 | - | 75.34 | 51.87 | 77.10 | 80.75 | 87.84 | 40.05 | 80.56 | 56.98 |
| DARR [24] | 69.77 | - | 74.74 | 53.77 | 72.31 | 73.24 | 94.08 | 54.18 | 89.90 | 45.96 |
| R50 U-Net [5] | 74.68 | 36.87 | 87.74 | 63.66 | 80.60 | 78.19 | 93.74 | 56.90 | 85.87 | 74.16 |
| U-Net [1] | 76.85 | 39.70 | 89.07 | 69.72 | 77.77 | 68.60 | 93.43 | 53.98 | 86.67 | 75.58 |
| R50 Att-UNet [5] | 75.57 | 36.97 | 55.92 | 63.91 | 79.20 | 72.71 | 93.56 | 49.37 | 87.19 | 74.95 |
| Att-UNet [23] | 77.77 | 36.02 | **89.55** | 68.88 | 77.98 | 71.11 | 93.57 | 58.04 | 87.30 | 75.75 |
| R50 ViT [5] | 71.29 | 32.87 | 73.73 | 55.13 | 75.80 | 72.20 | 91.51 | 45.99 | 81.99 | 73.95 |
| TransUNet [5] | 77.48 | 31.69 | 87.23 | 63.13 | 81.87 | 77.02 | 94.08 | 55.86 | 85.08 | 75.62 |
| SwinUNet | 79.13 | 21.55 | 85.47 | 66.53 | 83.28 | 79.61 | 94.29 | 56.58 | **90.66** | 76.60 |
| FocalUNet | **80.81** | **20.66** | 85.74 | **71.37** | **85.23** | **82.99** | **94.38** | **59.34** | 88.49 | **78.94** |

### 4.2 Implementation details

The Focal-UNet is implemented using Python 3.7 and Pytorch 1.8.1. We increase the data diversity by augmentation, using image flips and rotations. The input image size is 224×224. We train our model on an Nvidia 1080Ti GPU with 11GB memory. We did not use any pre-trained weights; the model is trained from scratch and initialized by the Kaiming method. During the training process, the batch size is 24, and the popular AdamW optimizer is used to optimize our model for backpropagation. The beta1, beta2, and the learning rate are set to be 0.5, 0.99, and 0.01, respectively. We also used the Cosine Annealing learning rate scheduler with warm restart, and the T_0 and T_mult are set to be the number of iterations and 2, respectively.

Table 2: Comparing the performance of Swin-UNet and Focal-UNet on the NeoPolyp dataset.

| Methods | Dice-Similarity coefficient (DSC) |
|---|---|
| Swin-UNet | 75.96 |
| Focal-UNet | **80.21** |

### 4.3 Experiment results on Synapse dataset

Table 1 compares the proposed Focal-UNet with previous state-of-the-art methods on the Synapse dataset. Because we used the same train and test sets setups as [6], therefore, we did not implement the methods [1], [5], [23-25], which were used in [6] for comparison, and the results are directly adapted from [6]. Experimental results depict that our Focal-UNet method achieves segmentation accuracy of 80.8% (DSC↑) and 20.66% (HD↓), which is the best performance among the previous state-of-the-art pure transformer-based methods. Furthermore, our method achieves better DSC and HD accuracy than the Swin-UNet method, demonstrating that increasing the network's receptive field using a focal module could help the UNet architecture handle the low and high-level (edges) features simultaneously. The segmentation results of our proposed method are compared visually with Swin-UNet in Figure 2. As we can see, the quality of our generated masks is higher than Swin-UNet, specifically on the edges. Due to the window partitioning procedure of the Swin blocks [4], the Swin-UNet also suffers from blockiness in its output which is solved by using the focal modulation procedure instead. As mentioned in [6], the convolution-based methods have an over-segmentation problem due to the locality of the convolution module. This problem is solved by balancing the receptive field of the UNet (i.e., the network considers local and global features simultaneously) architecture using focal modules.

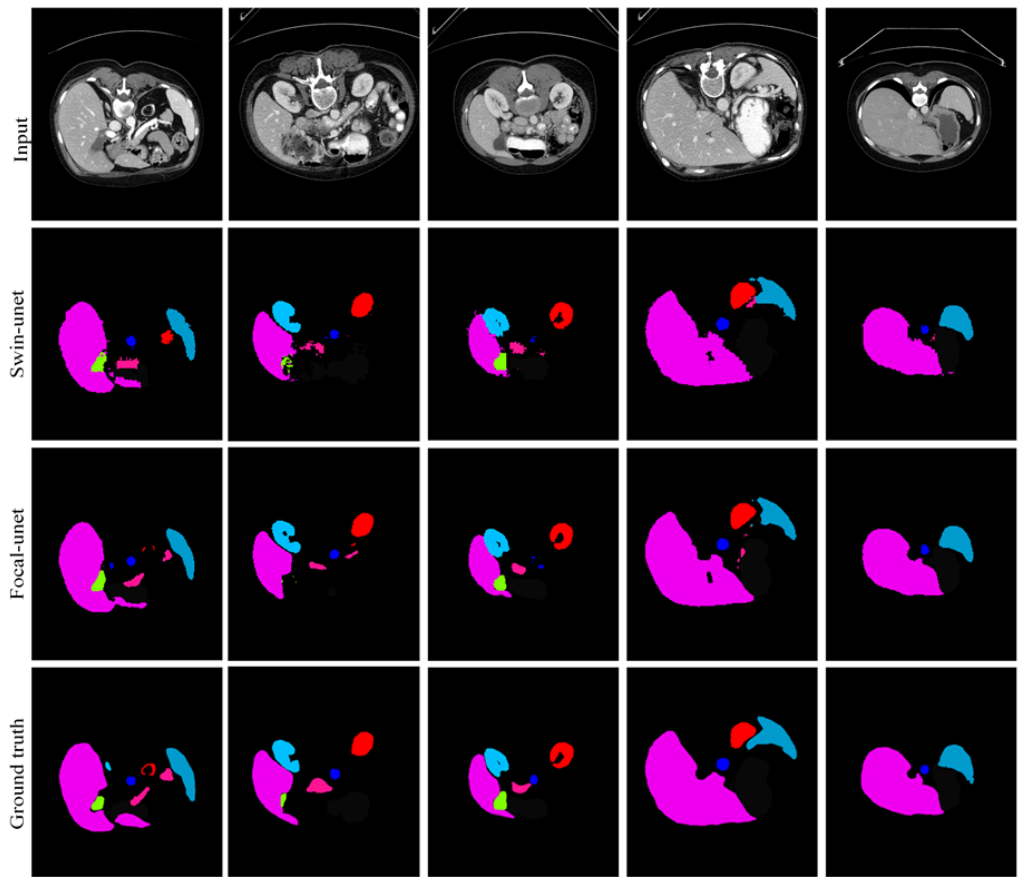

Figure 2: The segmentation results of Swin-UNet and Focal-UNet on Synapse dataset.

In this work, we demonstrate that by integrating a focal module with a U-shaped architecture with skip connections, the pure focal approach can better learn both local and long-range semantic information interactions, resulting in better segmentation results.

### 4.4   Experiment results on the NeoPolyp dataset

In this experiment, we showed the generalization of our proposed method for the medical image segmentation task. The Swin-UNet needs not only pre-trained weights but also a large training dataset. We trained our proposed approach and Swin-UNet on the NeoPolyp dataset to illustrate the capacity of Focal-UNet in working with a small dataset, which is very important in medical image tasks. Table 2 shows the comparison of the Focal-UNet with the Swin-UNet. As we can see, our proposed method achieves a 5% higher segmentation accuracy than Swin-UNet. For better comparison, we can see the outputs of Swin-UNet and Focal-UNet in Figure 3, demonstrating the quality of segmented regions. The ground truths of the test images are not publicly available; therefore, this figure does not include ground truth images.

## 5   Conclusion

We proposed a new focal-based U-shaped architecture for medical image segmentation in this work. As a result, our Focal-UNet model achieved better results than the state-of-the-art methods on the Synapse dataset. Also, our experiments demonstrate that the proposed Focal-UNet has excellent performance and generalization ability compared to the Swin-UNet for small training sets.

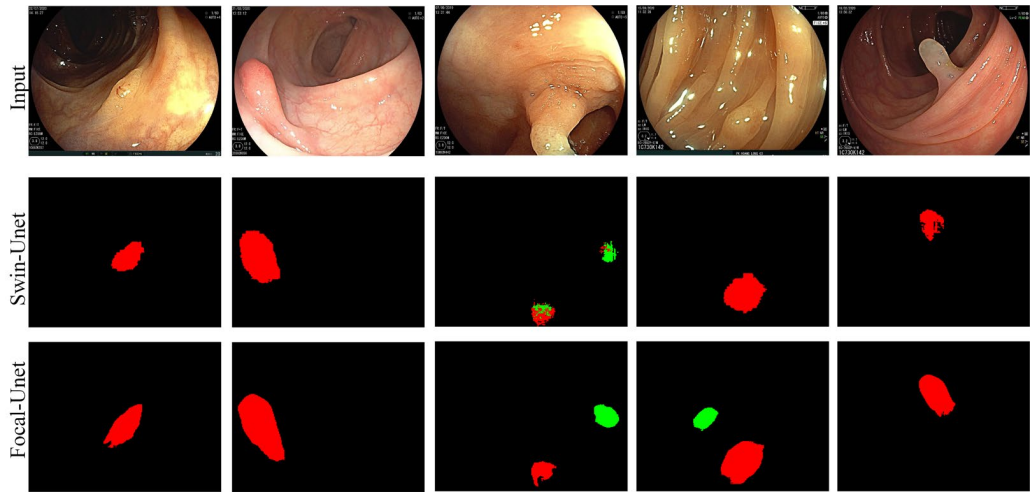

Figure 3: The segmentation results of Swin-UNet and Focal-UNet on NeoPolyp dataset.

**Future works:** The proposed Focal-UNet can model the problem space with a limited number of training data. This model could be used as the base network to construct more complex architectures, such as generative models with discrimination processes. It also could be applied for other medical purposes with a limited number of training data.

**References**

[1] O. Ronneberger, P. Fischer, and T. Brox, "U-net: Convolutional networks for biomedical image segmentation," in International Conference on Medical image computing and computer-assisted intervention, 2015, pp. 234–241.

[2] X. Xiao, S. Lian, Z. Luo, and S. Li, "Weighted Res-UNet for High-Quality Retina Vessel Segmentation," Proc. - 9th Int. Conf. Inf. Technol. Med. Educ. ITME 2018, pp. 327–331, 2018, DOI: 10.1109/ITME.2018.00080.

[3] A. Dosovitskiy et al., "An image is worth 16x16 words: Transformers for image recognition at scale," arXiv Prepr. arXiv2010.11929, 2020.

[4] Z. Liu et al., "Swin transformer: Hierarchical vision transformer using shifted windows," in Proceedings of the IEEE/CVF International Conference on Computer Vision, 2021, pp. 10012–10022.

[5] J. Chen et al., "TransUNet: Transformers make strong encoders for medical image segmentation," arXiv Prepr. arXiv2102.04306, 2021.

[6] H. Cao et al., "Swin-UNet: UNet-like pure transformer for medical image segmentation," arXiv Prepr. arXiv2105.05537, 2021.

[7] Z. Zhou, M. M. Rahman Siddiquee, N. Tajbakhsh, and J. Liang, "UNet++: A nested u-net architecture for medical image segmentation," in Deep learning in medical image analysis and multimodal learning for clinical decision support, Springer, 2018, pp. 3-11.

[8] H. Huang et al., "UNet 3+: A full-scale connected UNet for medical image segmentation," in ICASSP 2020-2020 IEEE International Conference on Acoustics, Speech and Signal Processing (ICASSP), 2020, pp. 1055–1059.

[9] X. Li, H. Chen, X. Qi, Q. Dou, C.-W. Fu, and P.-A. Heng, "H-DenseUNet: hybrid densely connected UNet for liver and tumor segmentation from CT volumes," IEEE Trans. Med. Imaging, vol. 37, no. 12, pp. 2663-2674, 2018.

[10] Z. Meng, Z. Fan, Z. Zhao, and F. Su, "ENS-UNet: End-to-end noise suppression U-Net for brain tumor segmentation," in 2018 40th Annual International Conference of the IEEE Engineering in Medicine and Biology Society (EMBC), 2018, pp. 5886-5889.

[11] J. Wu, E. Z. Chen, R. Rong, X. Li, D. Xu, and H. Jiang, "Skin lesion segmentation with C-UNet,"

in 2019 41st Annual International Conference of the IEEE Engineering in Medicine and Biology Society (EMBC), 2019, pp. 2785–2788.

[12] A. Vaswani et al., "Attention is all you need," Adv. Neural Inf. Process. Syst., vol. 30, 2017.

[13] H. Touvron, M. Cord, M. Douze, F. Massa, A. Sablayrolles, and H. Jégou, "Training data-efficient image transformers & distillation through attention," in International Conference on Machine Learning, 2021, pp. 10347–10357.

[14] J. Yang et al., "Focal self-attention for local-global interactions in vision transformers," arXiv Prepr. arXiv2107.00641, 2021.

[15] J. Yang, C. Li, and J. Gao, "Focal Modulation Networks," arXiv Prepr. arXiv2203.11926, 2022.

[16] H. Kan et al., "ITUNet: Integration Of Transformers And UNet For Organs-At-Risk Segmentation," in 2022 44th Annual International Conference of the IEEE Engineering in Medicine & Biology Society (EMBC), 2022, pp. 2123–2127.

[17] L. Qiao et al., "FcTC-UNet: Fine-grained Combination of Transformer and CNN for Thoracic Organs Segmentation," in 2022 44th Annual International Conference of the IEEE Engineering in Medicine & Biology Society (EMBC), 2022, pp. 4749-4753.

[18] C. Luo, J. Zhang, X. Chen, Y. Tang, X. Weng, and F. Xu, "UCATR: Based on CNN and Transformer Encoding and Cross-Attention Decoding for Lesion Segmentation of Acute Ischemic Stroke in Non-contrast Computed Tomography Images," in 2021 43rd Annual International Conference of the IEEE Engineering in Medicine & Biology Society (EMBC), 2021, pp. 3565-3568.

[19] J. M. J. Valanarasu, P. Oza, I. Hacihaliloglu, and V. M. Patel, "Medical transformer: Gated axial-attention for medical image segmentation," in International Conference on Medical Image Computing and Computer-Assisted Intervention, 2021, pp. 36–46.

[20] Y. Zhang, H. Liu, and Q. Hu, "Transfuse: Fusing transformers and cnns for medical image segmentation," in International Conference on Medical Image Computing and Computer-Assisted Intervention, 2021, pp. 14-24.

[21] G. P. Chen et al., "Asymmetric U-shaped network with hybrid attention mechanism for kidney ultrasound images segmentation." Expert Systems with Applications, 2023, p.118847.

[22] P. Ngoc Lan et al., "NeoUNet: Towards accurate colon polyp segmentation and neoplasm detection," in International Symposium on Visual Computing, 2021, pp. 15-28.

[23] O. Oktay et al., "Attention u-net: Learning where to look for the pancreas," arXiv Prepr. arXiv1804.03999, 2018.

[24] S. Fu et al., "Domain adaptive relational reasoning for 3d multi-organ segmentation," in International Conference on Medical Image Computing and Computer-Assisted Intervention, 2020, pp. 656-666.

[25] F. Milletari, N. Navab, and S.-A. Ahmadi, "V-net: Fully convolutional neural networks for volumetric medical image segmentation," in 2016 fourth international conference on 3D vision (3DV), 2016, pp. 565-571.
8